\documentclass[11pt,a4paper]{article}
\pdfoutput=1

\usepackage{jcappub}

\usepackage{bm}
\RequirePackage{color}

\makeatletter
\newlength{\apb@width}
\newcommand{\autoparbox}[2][c]{\settowidth{\apb@width}{#2}\parbox[#1]{\apb@width}{#2}}

\makeatother

\numberwithin{equation}{section}

\def\beq{\begin{eqnarray}}
\def\eeq{\end{eqnarray}}

\def\bea{\begin{eqnarray}}
\def\eea{\end{eqnarray}}

\def\0{{\boldsymbol 0}}

\DeclareRobustCommand{\SkipTocEntry}[4]{}

\begin{document}

\begin{titlepage}

\setcounter{page}{1} \baselineskip=15.5pt \thispagestyle{empty}
\bigskip\

\vspace{1cm}
\begin{center}
    
{\fontsize{20}{28}\selectfont  \sffamily \bfseries
The Self-consistent Matter Coupling of a Class of Minimally Modified Gravity Theories}
\end{center}

\vspace{0.2cm}

\begin{center}
{\fontsize{14}{30}\selectfont  Chunshan Lin}
\end{center}

\begin{center}
\textsl{Institute of Theoretical Physics, Faculty of Physics,
University of Warsaw, ul. Pasteura 5, Warsaw, Poland}
\end{center}

\vspace{1.2cm}
\hrule \vspace{0.3cm}
\noindent {\sffamily \bfseries Abstract} \\[0.1cm]
The self-consistent matter coupling is found in a broad class of minimally modified gravity theories which was discovered recently. All constraints in the theories remain first class and thus a graviton has only 2 local degrees of freedom. The cosmological solution of one of the examples in this class, the so-called square root gravity, exhibits a singularity freeness at high energy limit. At low energy limit, the theory smoothly connects to GR. A general feature of the theories in this class, with the self-consistent matter coupling discovered in our current work, is the non-trivial interaction among different components of matter sector.  We have also checked the Hamiltonian structure of a scalar QED coupling to the square root gravity in the same manner. All constraints in the theory are first class too and thus the local $U(1)$ gauge symmetry in scalar QED is preserved. 
\
\vskip 10pt
\hrule

\vspace{0.6cm}
 \end{titlepage}

 \tableofcontents

\newpage

\section{introduction}
 Is GR unique? According to Lovelock theorem \cite{lovelock1}\cite{lovelock2}, the answer is yes if we assume that (1) the space time is 3+1 dimensional; (2) metricity; (3) space-time diffeomorphism invariance; (4) at most second order derivative in the equation of motion. In terms of  the language of Hamiltonian analysis, the local gauge symmetry of GR, i.e. the space-time diffeomorphism invariance gives rise to 8 constraints, including 4 primary ones and 4 secondary ones, which are all first class. Therefore, a graviton in GR has only 2 polarizations. It is very intriguing to ask whether GR is still the unique effective description for  massless spin-2 particles if we demand all constraints in the theory to be first class in the first place. Or in other words, is there any theory that is as good as GR in the sense that all of constraints are first class, and therefore the structure of the theory at low energy scale is stable against quantum corrections? This problem was formulated and investigated in Ref. \cite{Lin:2017oow}, where a new class of minimally modified gravity theories was discovered. One of the most interesting examples is the so-called square root gravity, of which Hamiltonian and Lagrangian have very perculiar square root structures. The theory is free from cosmological singularity if matter sector couples to gravity in the minimal manner. However, as pointed out in Ref. \cite{Carballo-Rubio:2018czn}, the minimal coupling between gravity and matter renders the algebra unclosed and the theory becomes inconsistent. This inconsistency arises from the discrepancy between the local gauge symmetries of gravity and matter.
The symmetry in matter sector is 4 dimensional space-time diffeomorphism invariance, if matter minimally couples to gravity. 
However, temporal diffeomorphism invariance is broken in the gravity sector. Instead, gravity sector has a different type of local gauge symmetry, which associated to the first class Hamiltonian constraint in the vacuum. This symmetry is explicitly broken if matter couples to gravity in the minimal manner. 

On the other hand,  the graviton scattering amplitude computation exhibits a perturbative equivalence to GR  which holds up to 5 point function level for all possible helicity configurations in a Minkowskian vaccum \cite{Carballo-Rubio:2018czn}. Nevertheless, it is still premature to claim an equivalence between the square root gravity and GR, given that the self-consistent matter coupling in the square root gravity is still missing. Even when the theory is completely equivalent to GR in the vacuum, the non-trivial and non-minimal coupling between matter field and gravity still yields non-standard predictions that differ from GR \cite{Aoki:2018zcv}\cite{Aoki:2018brq}. This point is actually not new. For instance, given Einstein-Hilbert action, we still consider the theory as a modified gravity theory if matter couples to gravity in the Brans-Dicke manner. In our current work, for the first time, we find the self-consistent matter coupling in the square root gravity. Interestingly the cosmological solution exhibits the singularity freeness at high energy limit. Therefore, the theory gives different predictions that differ from GR even at the background level of Einstein equation.  It is straightforward to generalise our analysis to a broader class of theories.

This paper is organised as follows: In the section \ref{1scalar}, starting from the simplest case of one single scalar field, we derive a self-consistent way that a single scalar field couples to the square root gravity. We will also briefly discuss its cosmology as well as its generalization in this section. In the section \ref{multifield} we discuss how to couple multi matter fields to the square root gravity. We conclude in the section \ref{C&D}.

\section{One Single Scalar Field} \label{1scalar}
\subsection{Hamilontian analysis}
We find it is more convenient to start from the Hamiltonian. In this section, we will firstly introduce one single scalar field $\phi$ to represent the matter sector. Generalising to multiple matter components is straightforward and will be discussed later. We adopt the ADM decomposition of the 4-dimentional space-time, 
\beq
ds^2=-N^2dt^2+h_{ij}\left(dx^i+N^idt\right)\left(dx^j+N^jdt\right).
\eeq
In terms of Hamiltonian language the gravitational field has 10 canonical pairs, $\left(h_{ij},\pi^{ij}\right),\left(N,\pi_N\right)$ and $\left(N^i,\pi_i\right),$ where  $\pi^{ij},~\pi_N,~\pi_i$ are the conjugate momenta of $h_{ij},~N$ and $N^i$ respectively. The matter sector contains one canonical pair $\left(\phi,\pi_\phi\right)$. The  total Hamiltonian can be written as 
\beq\label{totalHam}
H=\int d^3x\left[N\mathcal{C}+N^i\mathcal{H}_i+\lambda_N\pi_N+\lambda^i\pi_i\right]\, ,
\eeq
where $\mathcal{C}\approx0$ is the Hamiltonian constraint, it can be splitted into the gravity part, which contains a perculiar square root structure \cite{Lin:2017oow}, and the matter part which is denoted as a generic function $\mathcal{U}$ of arguments $\frac{\pi_\phi^2}{h},\nabla_i\phi\nabla^i\phi,$ and $\phi$,
\beq\label{ham1}
\mathcal{C}\equiv -\xi_g\sqrt{h}B^{1/2}\left[CR+D-\frac{4}{Ah}\left(\pi^{ij}\pi_{ij}-\frac{1}{2}\pi^2\right)\right]^{1/2}
+\sqrt{h}~\mathcal{U}\left(\frac{\pi_\phi^2}{h},\nabla_i\phi\nabla^i\phi,\phi\right),
\eeq
where $\xi_g=\pm1$, $R$ is 3-d Ricci scalar, $\pi_{\phi}$ is the conjugate momentum of scalar field $\phi$.  The explicit form of $\mathcal{U}$ will be decided by the consistency condition of the theory. The coefficients $A,B,C$ and $D$ can be some functions of time, and their time dependences have to satisfy a consistency condition that we will discuss later.   We demand that $BD>0$ so that the Hamiltonian is real at the ground state of gravity. $\mathcal{H}_i\approx0$ are three momentum constraints, 
\beq
\mathcal{H}_i &\equiv&  -2 \sqrt{h}\nabla_j\left(\frac{\pi^j_{~i}}{\sqrt{h}}\right)+\pi_\phi\nabla_i\phi\,,
\eeq
they generate 3-d diffeomorphism, and  $\pi_N\approx\pi_i\approx0$ are four primary constraints.

With total Hamiltonian written in the form of eq. (\ref{totalHam}), it is obvious that the constraints $\pi_N\approx0,$ and $\pi_i\approx0$ are first class. While momentum constraint $\mathcal{H}_i\approx0$ is also first class because of the spatial diffeomorphism invariance of the theory \cite{Saitou:2016lvb}\cite{Lin:2017utd}.
The self-consistency of the theory requires that $\mathcal{C}\approx0$ must be first class as well, otherwise dimension of the phase space would be odd. Let us count the degrees as follows:  at each point there are 11 canonical pairs (including the one from matter sector) and thus 22 degrees in the phase space. The Hamiltonian constraint $\mathcal{C}$, if it is second class, eliminates one degree in the phase space. On the other hand,  the 7 first class constraints mentioned above eliminate 14 degrees in the phase space. At the end we have $22-1-14=7$ degrees in the phase space. The odd number of dimension in the phase space implies that the theory is inconsistent. Therefore Hamiltonian constraint must be first class. With $\mathcal{C}\approx0$ being first class we only need that 
\beq\label{key1}
\{\mathcal{C}[\alpha],\mathcal{C}[\beta]\}\approx0,
\eeq
where we have introduced the notation $\mathcal{O}[\alpha]\equiv \int d^3x \alpha\mathcal{O}$ for convenience. The direct computation of Poisson bracket eq. (\ref{key1}) gives us the following result, 
\beq\label{cond1}
\{\mathcal{C}[\alpha],\mathcal{C}[\beta]\}\approx \int \left(\beta\nabla^i\alpha-\alpha\nabla^i\beta\right)\cdot
\left[\frac{2B^2C}{A\cdot\mathcal{U}^2}\sqrt{h}\nabla_j\left(\frac{\pi^j_{~i}}{\sqrt{h}}\right)-\frac{\partial\mathcal{U}}{\partial Q}\frac{\partial\mathcal{U}}{\partial X}\pi_{\phi}\nabla_i\phi\right],
\eeq
where we have defined the following two variables to avoid clumsy notations, 
\beq\label{QX}
Q&\equiv& \frac{\pi_\phi^2}{2h},\qquad X\equiv \frac{1}{2}\nabla_i\phi\nabla^i\phi\,.
\eeq
The Poisson bracket eq. (\ref{cond1}) vanishes weakly if
\beq
\frac{B^2C}{A\cdot\mathcal{U}^2}=\frac{\partial\mathcal{U}}{\partial Q}\frac{\partial\mathcal{U}}{\partial X}\,.
\eeq
After rearranging the variables, the above equation can be rewritten as,
\beq\label{crit1}
\frac{\partial\tilde{\mathcal{U}}}{\partial Q}\frac{\partial\tilde{\mathcal{U}}}{\partial X}=1\,,
\eeq
where $\tilde{\mathcal{U}}\equiv\frac{1}{2}\sqrt{\frac{A}{B^2C}}\mathcal{U}^2$ and we have assumed that $\frac{A}{C}>0$. If $\frac{A}{C}<0$  a gradient instability appears even at low energy scale and therefore we are not interested in this case. The eq. (\ref{crit1}) has the following linear solution,
\beq
\tilde{\mathcal{U}}=\frac{1}{2}\Lambda+\zeta Q+\frac{1}{\zeta}X\,,
\eeq
where $\Lambda$, and $\zeta$  are generic functions of scalar field $\phi$ and time $t$, and the factor $1/2$ in front of $\Lambda$ is introduced for the later convenience.  Therefore we have
\beq
\mathcal{U}=\xi_m\left(\frac{B^2C}{A}\right)^{1/4}\sqrt{\zeta\frac{\pi_\phi^2}{h}+\frac{1}{\zeta}\nabla_i\phi\nabla^i\phi+\Lambda}\,,
\eeq
where $\xi_m\equiv\pm1$, i.e. there are positive and negative branches of solution. Interestingly, matter sector contains a similar square root structure.  The Hamiltonian constraint eq. (\ref{ham1}) requires that the Hamiltonian of gravity part and matter part summed up to zero. It implies that $\xi_g=\xi_m$ and therefore we will omit the subscripts of them.  We also need to take into account the possible explicit time dependences of the parameters in the theory. Firstly, please notice that both of gravity Hamiltonian and matter Hamiltonian contain $\sqrt{B}$ and thus the time dependence of $\sqrt{B}$, if any, can be absorbed into a redefinition of the lapse $N$.  Therefore, we will only consider $B=\text{constant}$ throughout this letter. The explicit time dependences of the rest of parameters in the Hamiltonian acquire the following consistency condition,
\beq
\frac{d\mathcal{C}}{dt}=\frac{\partial\mathcal{C}}{\partial t}+\{\mathcal{C},H\}\approx\frac{\partial\mathcal{C}}{\partial t}\approx0\,.
\eeq
This condition gives us 
\beq\label{cond2}
\frac{\partial\zeta}{\partial t}=0, \qquad\frac{\partial(AC)}{\partial t}=0,\qquad
\frac{\partial\Lambda}{\partial t}=\frac{\partial\left(D\sqrt{A/C}\right)}{\partial t}.
\eeq
No tertiary constraint is generated, and all 8 constraints, including Hamiltonian constraint, in the theory are first class, provided the coefficients in the theory satisfy the condition eq. (\ref{cond2}). On the other hand,  we can also introduce the scalar field $\phi$ dependence to coefficients $A, ~C$ and $D$ (the $\phi$ dependence of $B$ can be absorbed into a redefinition of the lapse $N$). The necessary condition of Hamilontian constraint being first class, i.e. eq. (\ref{cond1}), requires that  $\partial (AC)/\partial\phi=0,$ while $D$ can be an arbitrary function of scalar field $\phi$. 

These 8 first class constraints eliminate 16 degrees and at the end we have $22-16=6$ degrees in the phase space, or in other word, 3 dynamical degrees of freedom in physical space-time. One of them is the scalar degree from matter sector, and the rest of 2 degrees are 2 polarisations of a massless graviton. We conclude that there are only 2 local degrees of freedom in the gravity sector.  As we have known, first class constraints correspond to local gauge symmetries in a theory. Even though sometimes these symmetries are not manifest and transparent. The structure of the theory at low energies is  expected to be protected by these symmetries and thus stable against quantum corrections.

When $B>0$ and $D>0$, we obtain the Lagrangian for our gravity and matter by means of a Legendre transformation, 
\beq\label{action}
\mathcal{L}&=&\sqrt{h}N\xi\left\{M^4\sqrt{ \left(1+\frac{c_1}{M^2}\mathcal{K}\right)\left(1+\frac{c_2}{M^2}R\right)}-\sqrt{\left[\sqrt{\frac{c_2}{c_1}}M^4-\frac{\zeta}{N^2}\left(\dot{\phi}-N^i\partial_i\phi\right)^2\right]\left(\frac{1}{\zeta}\nabla_i\phi\nabla^i\phi+\Lambda\right)}\right\}.\nonumber\\
\eeq
where $M\equiv\left(BD\right)^{1/8},~ c_1\equiv \frac{AM^2}{B}\,,$  $c_2\equiv \frac{M^2C}{D}\,,$ and $\mathcal{K}\equiv K_{ij}K^{ij}-K^2$. 
We have to bear in mind that coefficients $c_1$, $c_2$,  $M$ and $\Lambda$ can be time dependent,  $\zeta$ and $\Lambda$ can be some generic functions of scalar field $\phi$.

Let us split $\Lambda$ into bare cosmological constant part and matter part, i.e. $\Lambda=\Lambda_0+2V(\phi),$  where $V(\phi)=0$ at the ground state of matter sector and a factor of 2 is introduced so that $V(\phi)$ plays the role of potential of scalar field in the weak field limit.  In the weak field limit, we expand the action as 
\beq
S=\int d^4x \sqrt{h}N\xi&&\left[M^4+\frac{M^2}{2}\left(c_1K^{ij}K_{ij}-c_1K^2+c_2R\right)
-M^2\Lambda_0^{1/2}\left(\frac{c_2}{c_1}\right)^{1/4}\right.\nonumber\\
&&\left.+\frac{\zeta\Lambda_0^{1/2}\left(\dot{\phi}-N^i\partial_i\phi\right)^2}{2N^2\left(c_2/c_1\right)^{1/4}M^2}-\frac{M^2\left(c_2/c_1\right)^{1/4}}{2\zeta\Lambda_0^{1/2}}\nabla_i\phi\nabla^i\phi-\frac{M^2(c_2/c_1)^{1/4}}{\sqrt{\Lambda_0}}V(\phi)+...\right],\nonumber\\
\eeq
 where ellipsis denotes the higher order terms.  At low energy scale, we recognize the effective Planck mass $M_p^2=\xi c_1M^2$, the sound speed of gravitational waves $c_g^2=\frac{c_2}{c_1}$, and the effective cosmological constant $\Lambda_{eff}=M^4-M^2\Lambda_0^{1/2}\left(c_2/c_1\right)^{1/4}$. We demand that $\xi c_1>0$ to ensure the ghost freeness at low energy scale. The sound speed of matter sector reads $c_s^2=M^4c_g/\Lambda_0\zeta^2$, and $\xi\zeta>0$ is required as the ghost free condition in the matter sector.  The difference between sound speeds of the gravitational waves and the scalar field is one of the evidences that our theory is different from GR. At low energy limit, GR and Lorentz invariance in the matter sector are recovered if we set 
\beq\label{recgr}
c_2=c_1,\qquad \xi\zeta=1,\qquad M^4=\Lambda_0.
\eeq
The above parameter choice also cancels out the cosmological constant term in the action, and therefore the space-time is Minkowskian at the ground state of matter sector. The deviation from GR appears only at high energy scale characterised by $M$. \\

\subsection{Cosmology} 
Now let us study the cosmology of the theory. We take the FLRW ansatz, 
\beq
ds^2=-N^2dt^2+a^2d\textbf{x}^2\,,
\eeq
 adopt the parameter choice in eq. (\ref{recgr}), and assume that $A, B, C, D$ and $c_1$ are all positive and independent of time. At background level, the Einstein equations read
\beq
3c_1M^2H^2&=&\frac{\frac{1}{2}\dot{\phi}^2+V(\phi)}{1+2V(\phi)/M^4}\,,\label{ein1}\\
2c_1M^2\dot{H}&=&\frac{-\dot{\phi}^2+\dot{\phi}^4/M^4}{1+2V(\phi)/M^4}\,.\label{ein2}
\eeq
It is easy to check that  eq. (\ref{ein1}) is consistent with eq. (\ref{ein2}) and thus Bianchi identity holds, provided that scalar field $\phi$ satisfies the equation of motion,
\beq
\left(1+\frac{2V}{M^4}\right)\ddot{\phi}+3H\left(1+\frac{2V}{M^4}\right)\left(1-\frac{\dot{\phi}^2}{M^4}\right)\dot{\phi}+\left(1-\frac{\dot{\phi}^2}{M^4}\right)\frac{\partial V}{\partial\phi}=0.
\eeq
A de-sitter attractor is spotted if we trace our universe back in time, where
\beq
\dot{\phi}^2\to M^4,\qquad \ddot{\phi}\to0,\qquad\dot{H}\to0,\qquad H^2\to M^2/6c_1,
\eeq
as $t\to -\infty$. 
The Hubble constant approaches to a constant value, instead of infinity in the limit $V(\phi)\to\infty$ in the far past.  Therefore, the Hawking-Penrose singularity theorem \cite{singularity} does not apply here and our theory is free from the cosmological singularity. 
The cosmic singularity freeness is another evidence that our theory differs from Einstein gravity with minimally coupled matter\footnote{If matter couples to gravity in a non-minimal Brans-Dickel manner, we can always recast the theory into Einstein gravity with minimally coupled matter content by a conformal transformation.}.

It is also important to examine the properties of cosmological perturbations. Firstly let's perturb our metric elements as follows,
\beq
N=1+\alpha, \qquad N_i=\partial_i\beta,\qquad h_{ij}=a^2e^{2\zeta}\left(\delta_{ij}+\gamma_{ij}\right),
\eeq
where $\alpha,\beta$ and $\zeta$ are scalar perturbation and $\gamma_{ij}$ is the tensor mode.  The residual symmetries allow us to omit the fluctuation of matter, as well as the scalar perturbations arising from the off-diagonal part of 3-d induced metric $h_{ij}$. Due to the SO(3) invariance of FLRW space-time, scalar modes and tensor modes decouple at linear perturbation level. After integrating out the non-dynamical degrees the quadratic action for scalar perturbation reads
\beq
S_{\zeta}=\int a^3\frac{\dot{\phi}^2}{2H^2}\left[\frac{1}{\Gamma}\cdot\dot{\zeta}^2-\Gamma\frac{k^2}{a^2}\zeta^2\right]
\eeq
where $\Gamma\equiv\sqrt{1-6c_1H^2/M^2}$. Noted that $\Gamma\to0$ when the space-time approaches to the de-sitter attractor in the far past. The coefficient of kinetic term of scalar mode diverges in this limit, which implies a weak coupling at high energy limit. This is in contrast to Einstein gravity whose couplings are strong at UV side. On the other hand, sound speed of scalar modes approaches to zero too. Similar properties can be found in tensor modes too. The quadratic action of tensor modes reads
\beq
S_{T}=\frac{c_1M^2}{8}\int a^3\left[\frac{1}{\Gamma}\dot{\gamma}_{ij}\dot{\gamma}_{ij}-\Gamma\frac{k^2}{a^2}\gamma_{ij}\gamma_{ij}\right].
\eeq
The coefficient of the kinetic term of gravitational waves diverges in the same way as scalar modes. All higher order couplings are highly suppressed by $\Gamma$ which approaches to zero at high energy density limit and then we end up with a free theory.

\subsection{Generalization to the whole class of MMG}
our analysis can be extended to the whole class of minimally modified gravity theories, whose Hamiltonian can be formally written as 
\beq\label{fofr}
\mathcal{C}=\sqrt{h}F\left(R+\lambda\Pi/h\right)+\sqrt{h}\mathcal{U}\left(Q,X,\phi\right)\,,
\eeq
where $\Pi\equiv \pi^{ij}\pi_{ij}-\frac{1}{2}\pi^2$, $Q$ and $X$ are defined in eq. (\ref{QX}). Let us take $F(x)\sim (x+\text{constant})^n$ as an example. The consistency condition eq. (\ref{cond1}) acquires that the Hamiltonian of matter sector must take the form 
\beq
\mathcal{U}\sim \left(\Lambda+\zeta Q+\frac{1}{\zeta}X\right)^n.
\eeq
We recover the results of square root gravity if $n=1/2$, and we recover GR if $n=1$. \\

\section{Multi Matter Fields}\label{multifield}
If matter contains multiple components, and then we introduce multi fields $\phi_I$ to represent each of components. By means of the same approach in the previous section, the condition \ref{key1} requires that the matter part of Hamiltonian should be rewritten as 
\beq\label{multimatter}
\mathcal{U}=\xi\left(\frac{B^2C}{A}\right)^{1/4}\sqrt{\sum_I\left(\zeta_I\frac{\pi_I^2}{h}+\frac{1}{\zeta_I}\nabla_i\phi_I\nabla^i\phi_I\right)+\Lambda}.~~ 
\eeq
Translating to Lagrangian, 
\beq
\mathcal{L}_m= -\xi \sqrt{h}N\left[\sqrt{\frac{c_2}{c_1}}M^4-\sum_I\frac{\zeta_I}{N^2}\left(\dot{\phi}_I-N^i\partial_i\phi_I\right)^2\right]^{1/2}\cdot\left(\sum_I\frac{1}{\zeta_I}\nabla_i\phi_I\nabla^i\phi_I+\Lambda\right)^{1/2}.
\eeq
A key feature to note in the above Hamiltonian and Lagrangian is the novel coupling among different components of the matter sector. The non-trivial interaction among different components of the matter sector is one of the general features of the theories in this class. 

We can also include matter fields with higher spin into the square root of eq. (\ref{multimatter}), following the same summation rule. For instance, let's investigate the Hamiltonian of a scalar QED couples to the square root gravity in the same manner as scalar fields do. The Hamiltonian of a scalar QED minimally couples to Einstein gravity is derived in the appendix \ref{app:sQED}. We can add a scalar QED into the total Hamiltonian eq. (\ref{ham1}) and couple it to square root gravity in the same manner,
\beq
\mathcal{U}=\xi\sqrt{h}\left(\frac{B^2C}{A}\right)^{1/4}&&\left[\frac{\pi_A^i\pi_A^jh_{ij}}{h}+\frac{1}{2}F_{ij}F^{ij}+\frac{\pi_1^2}{h}+\frac{\pi_2^2}{h}+2eA_i\left(\phi_1\nabla^i\phi_2-\phi_2\nabla^i\phi_1\right)+\nabla_i\phi_1\nabla^i\phi_1\right.\nonumber\\
&&\left. +\nabla_i\phi_2\nabla^i\phi_2+\left(m^2+e^2A_iA^i\right)\left(\phi_1^2+\phi_2^2\right)+\frac{\pi_\chi^2}{h}+\nabla_i\chi\nabla^i\chi+V(\chi)+\Lambda\right]^{1/2},\nonumber\\
\eeq
where the definition of each of notations can be found in the Appendix \ref{app:sQED}. As a toy model, we have introduced an additional scalar field $\chi$ to represent the rest of part of the world. There are 10 constraints in the theory,  8 of them are of gravity part and 2 of them are of QED part. They are 
\beq
\mathcal{C}&\approx&0,\qquad\mathcal{H}_i\approx0,\qquad\pi_N\approx0,\qquad\pi_i\approx0,\nonumber\\
\pi_A^0&\approx&0,\qquad \mathcal{G}\equiv -\partial_i\pi_A^i+e\left(\pi_1\phi_2-\pi_2\phi_1\right)\approx0,
\eeq
where $\pi_A^0$ is the conjugate momentum of $A_0$, and $\mathcal{G}\approx0$ is the Gauss law.  By a straight forward computation we find that Poisson brackets of any two constraints in the theory are vanishing. Therefore, all constraints are first class. The $U(1)$ gauge symmetry is preserved because the 2 constraints in the gauge field sector are first class. We expect that we can self consistently couple the whole standard model of particle physics to our square root gravity in this manner, without spoiling the gauge symmetries of standard model.

\section{Conclusion and Discussion} \label{C&D}
The self-consistent matter coupling of a broad class of minimally modified gravity theories has been found.  The existence of this broad class of theories, as well as the self-consistent matter coupling discovered in our current work, may have challenged the distinctive role of GR as the unique non-linear theory for massless spin-2 particles. However, the perturbative equivalence between this class of theories and GR (up to 5 point function level in a Minkowskian vacuum) was found in Ref. \cite{Carballo-Rubio:2018czn}. It suggests that this class of theories is probably just GR in a different guise. If this is indeed the case, our current work can be understood as some novel type of matter coupling in GR, which was overlooked in the past. We can categorize our theories into type-I minimally modified gravity theories  \cite{Aoki:2018brq}: the gravity theories are equivalent to GR in the vacuum, but non-trivial matter coupling gives predictions differ from that of GR.  The Brans-Dicke theory, as well as another example that matter couples to gravity via a canonical transformation \cite{Aoki:2018zcv}\cite{Aoki:2018brq}, are two examples of this kind. 

The cosmology of the square root gravity exhibits an appealing property of cosmological singularity freeness. However, the singularity of spherical static solution still exists. One of the possible way to kill this singularity is to introduce a $\phi$ dependence of $D$ in the Hamiltonian eq. (\ref{ham1}). Noted that $D$ has dimension $[\text{mass}]^4$, therefore we can choose a function dependence of $D$ on $\phi$ so that it relates to the local energy density of matter at high density limit. The effective Newtonian constant is proportional  to $D^{-1/2}$ and thus gravity becomes weaker at short distances and high density regions. In a model of collapsing of an overdense region, the collapsing comes to a halt after the local matter density reaches a critical value. This may offer us some new thoughts to attack the problem of quantum gravity. \\

\section*{Acknowledgement}
 We would like to thank  Katsuki Aoki, Guillem Domenech, Hanna Lin, and Shinji Mukohyama for the useful discussions.  This work is carried out under POLONEZ programme of Polish National Science Centre, No.~UMO-2016/23/P/ST2/04240,  which has received funding from the European Union's Horizon 2020 research and innovation programme under the Marie Sklodowska-Curie grant agreement No.~665778.

\begin{appendix}

\section{Hamiltonian Analysis of a Scalar QED minimally couples to Einstein gravity}\label{app:sQED}

The action of a scalar QED minimally couples to Einstein gravity, 
\beq
S=\int d^4x\sqrt{-g}\left[\frac{M_p^2}{2}\mathcal{R}-\frac{1}{4}F_{\mu\nu}F^{\mu\nu}-\frac{1}{2}\left(\mathcal{D}_{\mu}\Phi\right)^\dagger\mathcal{D}_{\mu}\Phi-\frac{1}{2}m^2\Phi^\dagger\Phi\right],
\eeq
where $\mathcal{D}_\mu\equiv \partial_\mu-ieA_\mu$ and $\Phi$ is a complex scalar $\Phi\equiv \phi_1+i\phi_2$. 
The conjugate momenta of gauge field and scalars read
\beq
\pi_A^0&\equiv& \frac{\partial\mathcal{L}}{\partial \dot{A}_0}\approx0,\nonumber\\
\pi_A^i&\equiv&\frac{\partial \mathcal{L}}{\partial \dot{A}_i}=\frac{\sqrt{h}}{N}\left(F_{0j}-N^kF_{kj}\right)h^{ij}, \nonumber\\
\pi_1&\equiv& \frac{\partial\mathcal{L}}{\partial\dot{\phi}_1}=\frac{\sqrt{h}}{N}\left[\left(\dot{\phi}_1-N^i\partial_i\phi_1\right)+e\phi_2\left(N^i\partial A_i-A_0\right)\right],\nonumber\\
\pi_2&\equiv& \frac{\partial\mathcal{L}}{\partial\dot{\phi}_2}=\frac{\sqrt{h}}{N}\left[\left(\dot{\phi}_2-N^i\partial_i\phi_2\right)-e\phi_1\left(N^i\partial A_i-A_0\right)\right],
\eeq
where we have adopted ADM formalism. 
The Hamiltonian of scalar QED thus is calculated as
\beq
H_{QED}=\int d^3x\sqrt{h}N&&\left[\frac{\pi_A^i\pi_A^jh_{ij}}{2h}+\frac{1}{4}F_{ij}F^{ij}+\frac{\pi_1^2}{2h}+\frac{\pi_2^2}{2h}+eA_i\left(\phi_1\nabla^i\phi_2-\phi\nabla^i\phi_1\right)+\frac{1}{2}\nabla_i\phi_1\nabla^i\phi_1\right.\nonumber\\
&&~\left.+\frac{1}{2}\nabla_i\phi_2\nabla^i\phi_2+\frac{1}{2}\left(m^2+e^2A_iA^i\right)\left(\phi_1^2+\phi_2^2\right)\right]\nonumber\\
&&+N^i\left[\pi_1\partial_i\phi_1+\pi_2\partial_i\phi_2-eA_i\left(\pi_1\phi_2-\pi_2\phi_1\right)+\pi^j_AF_{ij}\right]\nonumber\\
&&+A_0\left(e\pi_1\phi_2-e\pi_2\phi_1-\partial_i\pi^i_A\right).
\eeq
Let's collect all of ingredients calculated above, and write down the total Hamiltonian, 
\beq
H_{tot}=\int d^3x\left[N\mathcal{C}+N^i\mathcal{H}_i+\lambda_N\pi_N+\lambda^i\pi_i+\lambda_{A}\pi_A^0+A_0\mathcal{G}\right]
\eeq
where $\mathcal{G}\equiv -\partial_i\pi_A^i+e\left(\pi_1\phi_2-\pi_2\phi_1\right)$ is the Gauss law ($\mathcal{G}$ stands for $Gauss$), $\mathcal{C}$ and $\mathcal{H}_i$ are Hamiltonian constraint and momentum constraints,
\beq
\mathcal{C}&\equiv&\sqrt{h}\left[\frac{2}{h}\left(\pi^{ij}\pi_{ij}-\frac{1}{2}\pi^2\right)-\frac{1}{2}R+\frac{\pi_A^i\pi_A^jh_{ij}}{2h}+\frac{1}{4}F_{ij}F^{ij}+\frac{\pi_1^2}{2h}+\frac{\pi_2^2}{2h}+\frac{1}{2}\nabla_i\phi_1\nabla^i\phi_1+\frac{1}{2}\nabla_i\phi_2\nabla^i\phi_2\right.\nonumber\\
&&\qquad\left. +eA_i\left(\phi_1\nabla^i\phi_2-\phi_2\nabla^i\phi_1\right)+\frac{1}{2}\left(m^2+e^2A_iA^i\right)\left(\phi_1^2+\phi_2^2\right)\right],\\
\mathcal{H}_i&=&-2\sqrt{h}\nabla_j\left(\frac{\pi^j_{~i}}{\sqrt{h}}\right)+\pi_1\partial_i\phi_1+\pi_2\partial_i\phi_2+\pi_A^jF_{ij}-eA_i\left(\pi_1\phi_2-\pi_2\phi_1\right)\,.\label{momcs}
\eeq

The momentum constraint eq. (\ref{momcs}) does not generate spatial diffeomorphism. The spatial diffeomorphism generator must be the combination 
\beq
\tilde{\mathcal{H}}_i&\equiv&\mathcal{H}_i+\mathcal{G}\cdot A_i\nonumber\\
&\approx&-2\sqrt{h}\nabla_j\left(\frac{\pi^j_{~i}}{\sqrt{h}}\right)+\pi_1\partial_i\phi_1+\pi_2\partial_i\phi_2+\pi_A^iF_{ij}-A_i\partial_j\pi_A^j\,,
\eeq
We can check that it does generate the spatial diffeomorphism. For instance,
\beq
\{A_i,~\xi^j\tilde{\mathcal{H}}_j\}\approx\mathcal{L}_\xi A_i=\xi^j\partial_jA_i+A_j\partial_i\xi^j.
\eeq
We have a whole set of constraints 
\beq
\Phi_A=\left(\mathcal{C},~\tilde{\mathcal{H}}_i,~\pi_N,~\pi^i,~\pi_A^0,~\mathcal{G}\right),
\eeq
where $\tilde{\mathcal{H}}_i\approx\pi_N\approx\pi^i\approx\pi_A^0\approx0$ are all obviously first class. We just need to check the Poisson brackets of the rest of two constraints $\mathcal{C}\approx0$ and $\mathcal{G}\approx0$, and we have 
\beq
\{\mathcal{G}[\alpha],\mathcal{C}[\beta]\}\approx0,\qquad \{\mathcal{G}[\alpha],\mathcal{C}[\beta]\}\approx0\,,\qquad\{\mathcal{C}[\alpha],\mathcal{C}[\beta]\}\approx0\,.
\eeq
Therefore, all constraints in the theory are first class, as we expect of course!

\end{appendix}

\end{document}